\documentstyle[12pt,aaspp4]{article}

\lefthead{Secker and Harris}
\righthead{The Coma Cluster Dwarf-to-Giant Ratio}

\begin{document}

\title{The Early-Type Dwarf-to-Giant Ratio and Substructure in the 
Coma Cluster}

\author{Jeff Secker\altaffilmark{1} and William E. Harris\altaffilmark{2}}

\affil{Department of Physics and Astronomy \\ McMaster University \\
Hamilton, Ontario L8S 4M1 \\ Canada}

\altaffiltext{1}{Present address: Washington State University, Program in 
Astronomy, Pullman, WA 99164-3113}

\altaffiltext{2}{Visiting Astronomer, KPNO, operated by AURA, Inc.\ under 
contract to the National Science Foundation.} 

\begin{abstract}
We have obtained new CCD photometry for a sample of $\simeq 800$
early-type galaxies (dwarf and giant ellipticals) in the central 700
arcmin$^2$ of the Coma cluster, complete in color and in magnitude to
$R = 22.5$ mag ($M_R \simeq -12$ mag for $H_0 = 86$ km/sec/Mpc).  The
composite luminosity function for all galaxies in the cluster core
(excluding NGC 4874 and NGC 4889) is modeled as the sum of a Gaussian
distribution for the giant galaxies and a Schechter function for the
dwarf elliptical galaxies.  We determine that the early-type
dwarf-to-giant ratio (EDGR) for Coma is identical to that measured for
the less rich Virgo cluster; i.e., the EDGR does not increase as
predicted by the EDGR-richness correlation.  We postulate that the
presence of substructure is an important factor in determining the
cluster's EDGR; that is, the EDGR for Coma is consistent with the Coma
cluster being built up from the merger of multiple less-rich galaxy
clusters.
\end{abstract}

\keywords{galaxies: clusters: individual (Coma) --- galaxies: formation ---
galaxies: evolution --- galaxies: luminosity function}

\section{INTRODUCTION} 

In their seminal study of dwarf galaxy populations in nearby groups,
Ferguson and Sandage \markcite{fs91} (1991, = FS91; Ferguson \&
Binggeli \markcite{fb94} 1994) defined the early-type dwarf-to-giant
ratio (EDGR) as the total number of early-type dwarf galaxies $N_{\rm
d}$ normalized to the number of early-type giant galaxies, $N_{\rm
g}$.  For the seven groups which they studied, they determined that
the EDGR is strongly correlated with the richness of the group, in the
sense that it increases monotonically with cluster richness, and
appears to be independent of distance from the cluster center.  This
correlation should provide insights into both the initial formation of
dE galaxies, and environmental influences on their subsequent
evolution.  An excellent test of the EDGR correlation is to extend the
analysis to much richer cluster environments than were originally
surveyed in the studies mentioned above.  Eventually, correlations
with other cluster properties (e.g., X-ray luminosity, velocity
dispersion, Bautz-Morgan class, substructure) can then be determined.

In their analysis, FS91\markcite{fs91} calculate the EDGR for four
different luminosity limits in the range $M_{B_T} = -15.5$ down to
$M_{B_T} = -12.5$.  In each case, $N_{\rm d}$ and $N_{\rm g}$
represent a sum of all early-type galaxies above that limiting
magnitude: with $N_{\rm g}$ constant for all cases, the EDGR increases
for fainter limiting magnitudes.  FS91\markcite{fs91} assigned
galaxies in several nearby groups and clusters into morphological
classes using visual inspection from photographic plates. By
definition, they assign $N_{\rm d}$ = N(dE+dE,N+dS0) and $N_{\rm g}$ =
N(E+S0).  In their analysis, the most convincing demonstration of the
EDGR-richness correlation was obtained by using $N_g$ itself as a
measure of the cluster richness. Then a plot of $N_{\rm d}$ versus
$N_{\rm g}$ shows clearly that the number of dwarfs rises faster than
the number of giants: the ratios $N_d/N_g$ for Fornax and Virgo (Abell
richness class 1, two of the richest clusters in their study) are
clearly larger than for Leo, Dorado and NGC 1400 (the poorest groups).

In a separate study of dwarf galaxies near early-type galaxies, Vader
and Sandage \markcite{vs91} (1991) confirmed the observed correlation
for the EDGR to increase with environment richness. They found that
giant galaxies in the richest groups are typically $\sim 2\times$
overabundant in companion dE galaxies when compared to the the mean
value of their sample.  In contrast, isolated field galaxies are
typically $\sim 10\times$ underabundant in dE companions relative to
the sample mean.  It is worthwhile to note here that the definition of
richness used by FS91\markcite{fs91} and Vader and Sandage
(1991\markcite{vs91}) differs from the Abell richness
class. Abell\markcite{a58} (1958) defined his ``richness class'' as a
number ranging from 0 to 5, corresponding to the number of counted
galaxies which are not more than two magnitudes fainter than the third
brightest member.  This scheme allows considerable variation in
numbers and densities within individual richness classes, and for
consistency, it requires imaging of the entire cluster field. The
FS91\markcite{fs91} definition of richness as equivalent to $\log
N_{\rm g}$ provides an alternative method for ordering groups by
richness, provided again that a considerable and consistent fraction
of the cluster is measured.

The Coma cluster is the nearest of the very rich Abell clusters (Abell
1656, richness class 2, $v \simeq 7000$ km/s). In this paper, we describe
new luminosity and color data for the galaxies in the cluster core
which allow us to define the EDGR accurately: in Section 2, we
summarize the observations and the derivation of the EDGR.  In Section
3, we discuss some implications of our data, particularly an
interpretation of EDGR-richness correlation in the context of cluster
merger and observed substructure.

\section{THE EARLY-TYPE DWARF-TO-GIANT RATIO}

Our new CCD photometry is described fully in Secker
(1995\markcite{s95a}=S95) and Secker \& Harris
(1996\markcite{s95b}=SH96). In brief, the KPNO 4m was used in April
1991 to mosaic-image the central region of the Coma cluster in two
colors ($B,R$), with the prime-focus camera and the TE 2K CCD
detector.  The total area covered was $\simeq 700$ arcmin$^2$, which
included a field centered on each of NGC 4874 and NGC 4889, and one
extending nearly 23 arcmin south of NGC 4874.  A control field located
about 2.1 degrees east of NGC 4874 was also imaged, for statistical
correction of the background population.  Total exposure times were
2700 seconds for both $R$ and $B$, with the seeing between 1.1 and 1.3
arcsec.  Since our targets were E and dE galaxies (characteristically
old stellar populations), we detected objects on the $R$-band images,
which had a slightly deeper limiting magnitude for these moderately
red objects. Total magnitudes for all detected objects on the frames
were determined by a curve-of-growth algorithm (i.e., apertures of
radius $2r_1$; see S95) and were tested for systematic correctness by
extensive Monte Carlo simulations of the images.  Analysis also showed
that detections of faint galaxies are $\gtrsim 80$ percent complete to
$R= 22.5$ mag: brighter that this limit, (minor) completeness
corrections are made to the measured luminosity function of the
detected objects; anything fainter than this limit was discarded so as
to keep only the highest-confidence data.  As well, we are complete in
$(B-R)$ to this same level.  Typical uncertainties near our
completeness limits are $\pm0.06$ mag in $R$ and $\pm0.12$ mag in
$(B-R)$.

Concerning our detection completeness of dwarf galaxies, we will
comment on the two cases involving spatially-extended giant galaxies,
near which our detection of dwarf galaxies is generally quite a bit
more incomplete.  In the end, we conclude that this spatial
incompleteness does not not strongly affect our computed EDGR. (i)
{\em Very near to brightest giants:} While a bias against detecting
close companion galaxies in a cluster environment such as Coma
certainly exists in wide-field data such as this, we note that the
majority of the dwarf galaxy population is not gravitationally bound
to any of the giant galaxies (Binggeli \markcite{b93} 1993; Ferguson
\markcite{f92} 1992; Vader \& Chaboyer \markcite{vc92} 1992). Thus, we 
do not expect this bias to affect our computed EDGR since the affected
area is quite small in total.  (ii) {\em Within an arcmin of the
supergiant galaxies NGC 4874 and NGC 4889:} As detected by SH96 and by
Bernstein et al. (1995), the radial number density profiles for dE
galaxies increases towards NGC 4874.  However, Thompson \& Gregory
(1993=TG93) determined that for their sample of bright dE galaxies,
this radial profile could be scaled downwards (in central surface
density), to match that of the early-type giant galaxies; this is a
direct indication that the EDGR is constant with position in the
cluster (provided that the limiting magnitude to which the dE galaxies
are counted is not too faint; see SH96).  This result of TG93 is
analogous to the result of FS91 for the Virgo cluster, that the EDGR
is independent of clustercentric radius.  Thus, the true EDGR for the
two excluded regions around the supergiant galaxies is consistent with
our computed value for the cluster core, within fluctuations due to
small number statistics arising from the small area concerned.  (Note
that these two incomplete regions comprise a mere $\simeq 1.6$ percent
of the total area studied).

The process of eliminating contaminating objects (foreground stars and
noncluster galaxies) was done in two parts.  Non-stellar objects were
reasonably well resolved in a parameter space of $r_{-2}$ (an image
moment which sensitively measures central concentration; see S95,
SH96) versus total magnitude $R$. The sequence of starlike objects in
this plane ($r_{-2} \lesssim 1.6$ pixels) brighter than $R = 19.5$ mag
is easily distinguished, and these bright starlike objects were
discarded. (This is only a minor correction, since the overwhelming
majority of detected objects in the Coma field are nonstellar, faint
galaxies.) Next, the $R$, $(B-R)$ color-magnitude diagram of all the
remaining objects was used to remove the non-cluster galaxies from the
sample.  In this diagram (Figure \ref{fig1}), it is clear that the
Coma galaxies populate a narrowly defined sequence, confined
(conservatively) to the range $0.7 \le (B-R) \le 1.9$ mag.  We
discarded any objects outside of this color range, leaving a total of
2526 candidate galaxies ($R \le 22.5$ mag) in the cluster core,
compared to 694 objects on the control field.  With a total area 2.57
times less than the cluster field area, the control-field number
density is $2.56\pm0.10$ objects arcmin$^{-2}$.

A comparison of our background number density with other similar
studies is worthwhile.  Prior to color discrimination, but within the
same magnitude range, our sample of control-field objects numbered
1146, or a number density of 4.23 objects arcmin$^{-2}$.  By
comparison, Bernstein et al. (1995) derive an $R$-band control-field
number density given by 5.46 objects arcmin$^{-2}$, over the same
magnitude range. (This number corresponds to an average over five
randomly located control fields, and not related to the position of
the Coma cluster.)  In his analysis of field galaxies, Koo (1986)
derived a mean galaxy number density of 3.69 arcmin$^{-2}$ to a
magnitude cutoff of $22.5$ mag in the $F$ filter (their red band pass)
in a field located near to (but outside of) the Coma cluster. Thus
prior to imposing the color constraints, our computed background
number density is comparable to other derived values.  This comparison
clearly illustrates the benefits to deriving colors for all detected
objects: the level of contamination by background objects is greatly
reduced, without affecting the population of true cluster galaxies.

\placefigure{fig1}

The expected luminosity range for all types of dwarf galaxies is
fainter than $R \simeq 15.5$ mag, and as illustrated in Figure
\ref{fig1}, these objects are generously included in our survey. We do
not attempt to divide these dwarf galaxies into their morphological
subgroups (nucleated and non-nucleated dEs, irregulars, etc.), simply
because we have insufficient spatial resolution.  However, the vast
majority of our sample must be dwarf ellipticals uncontaminated by
late-type systems: in their analysis of late-type dwarf (dIr) galaxies
in Coma, TG93 find a significant decrease in the number density of dIr
galaxies in the Coma cluster core, with typical values of 14 -- 40 dIr
per square degree to $R \lesssim 18.6$.  Scaling these densities to
our 0.194 square degree field, we therefore expect only 3 -- 8 dIrs to
this limit, considerably less than the number of dE galaxies
(cf. Table \ref{tbl-2}).  Fainter than this, the luminosity function
for dIr galaxies is bounded (Binggeli, Sandage \& Tammann
\markcite{bst88} 1988), unlike the still-rising luminosity function
for dE galaxies.

Brighter than $R \simeq 15.5$, our sample includes all detected giant
galaxies, excluding only the supergiants NGC 4874 and NGC 4889.  While
the number of late-type {\it giants} in the cluster core is small, it
is not negligible.  Thompson \& Gregory \markcite{tg80} (1980) detect
a total of 34 spiral galaxies in a 1.49 square degree region of the
cluster core, such that in our sample we would expect 2 -- 7 spirals.
However, since we cannot individually separate and discard the
late-type dwarf galaxies, we do not discard the few late-type giants
either. Thus we are assuming that the contaminating late-type galaxies
compose similar fractions of the bright and faint galaxy
populations. While this late-type dwarf-to-giant ratio is {\em
slightly less} than the EDGR for rich clusters (FS91\markcite{fs91}),
the effect of sample contamination by late-type galaxies on the
computed EDGR will be small when averaged in with the much larger
sample of early-type galaxies.

In both previous analyses, FS91\markcite{fs91} and TG93\markcite{tg93} used
visual classification to separate galaxies into their morphological
subgroups, with surface brightness as the primary discriminator
between giants and dwarfs.  TG93, and more recently Biviano et
al. (1995\markcite{b95}) determined that the composite luminosity
function (LF) for the Coma cluster is well fit by a Gaussian
distribution for the bright cluster members and a Schechter function
for the dwarf galaxies. The sum of these two functions, with the
appropriate normalization, has been found to provide a better
representation of the total luminosity function than a single
Schechter function.  Conveniently, the resulting scaled functions
provide the {\em relative contribution to the LF by each of the dwarf
and giant galaxy populations}. It is this technique which we adopt in
principle to determine the EDGR value for our sample of cluster
galaxies.

The net galaxy LF which we analyze here is plotted in Figure
\ref{fig2}, where the number of galaxies per 0.5-mag bin is
represented by the solid circle and the associated error bars.  Both
the raw galaxy LF and the background field LF were first corrected for
small incompleteness effects using a completeness function derived
from the overlapping regions of the three adjacent Coma fields (i.e.,
the sample is essentially 100 percent complete to 18th mag, falling
gradually to 80 percent complete at 22.5 mag). The error bars for the
net LF plotted in Figure \ref{fig2} represent the Poisson errors from
the two individual LFs added in quadrature.

Weighted least-squares fits of a Gaussian plus a Schechter function to
the luminosity function of Figure \ref{fig2} then allowed us to solve
for the relative contribution from each of the giant and dwarf galaxy
populations.  This fit formally involves six parameters: the peak
magnitude $m^0$, dispersion $\sigma$ and normalization $\kappa_1$ of
the Gaussian function for the giants, and the characteristic magnitude
$m^{\star}$, the faint-end slope $\alpha$ and normalization $\kappa_2$
of the Schechter LF for the dwarf galaxies.  However, two of these
parameters ($\sigma, \alpha$) can be reasonably constrained {\em a
priori}.  A separate analysis of the faint-end slope of our LF yields
$\alpha = -1.41\pm0.05$ (SH96, and consistent with TG93 and Bernstein
et al.  1995), and we limit the Gaussian dispersion to lie within the
realistic range $\sigma = 1.0\pm0.2$ mag (see TG93\markcite{tg93} and
Biviano et al. 1995\markcite{b95}).  The numerical results
corresponding to nine separate cases are summarized in Table
\ref{tbl-1}; on the basis of the reduced $\chi^2$ values, any of these
can be accepted on statistical grounds.  The model fits corresponding
to extreme values for $\sigma$ and $\alpha$ will define upper and
lower limits to the EDGR, while the adopted EDGR represents the
average of these extreme values.  For all models considered here, the
maximum EDGR is for the smallest $\sigma$ combined with a steep
faint-end slope (i.e., fit 1a), while the minimum EDGR is for the
largest $\sigma$ combined with a shallow faint-end slope (i.e., fit
5b).

\placetable{tbl-1}

As mentioned above, the calculation of the EDGR involves numerically
integrating the Schechter function to an arbitrary limiting magnitude.
To remain consistent with the limiting magnitudes used by
FS91\markcite{fs91}, it is necessary to convert our apparent $R$ total
magnitudes to their absolute total magnitudes $M_{B_T}$. To do so, we
adopt the same parameters used by TG93\markcite{tg93}: a difference in
distance moduli (Coma--Virgo) of 3.68 mag and $B$-band $K$-corrections of
0.12 mag for Coma and 0.02 mag for Virgo.  To this we add the Virgo
distance modulus of 31.7 used by FS91\markcite{fs91}\footnote{This is
not the Virgo distance modulus which we would adopt; its only purpose
here is to transform their absolute magnitudes back to appropriate
{\it apparent} magnitudes at the Coma distance.}.  Finally, we adopt a
mean dE color of $(B-R) = 1.4$ mag for $R \simeq 18.0$
(S95\markcite{s95a}; SH96\markcite{s95b}). This leads to a conversion
factor, from our $R$ mag to a $M_{B_T}$ mag on the FS91 scale, of
$M_{B_T} \simeq R - 34.1$.  Then the four limiting magnitudes which we
adopt are $R = 18.6$, 19.6, 20.6 and 21.6 mag, corresponding to their
$M_{B_T} = -15.5$, --14.5, --13.5 and --12.5 mag limits.

We integrate the Gaussian function over the interval $12.5 \leq R \leq
18.5$ mag, and obtain values for the number of giants of N(E+S0) =
30.51, 28.46, 31.95, 33.96, 36.72, 39.11, 41.42, 40.44 and 42.20 for
model fits 1, 1a, 1b, 2, 3, 4, 5, 5a and 5b respectively. In Table
\ref{tbl-2} we provide the corresponding number of dwarfs
N(dE+dE,N+dS0) and the Coma EDGR values, obtained by numerically
integrating the Schechter function over the appropriate magnitude
ranges, using the parameter values relevant to the five different
fits.  In Figure \ref{fig2}, we illustrate one of the model fits
(i.e., fit 3, with $\sigma = 1.0$ mag and $\alpha = -1.41$) to the
composite luminosity function.  The dotted lines designate the
individual contributions from the dwarfs and giants, while the solid
line is the direct sum of the two.

\placetable{tbl-2}
\placefigure{fig2}

Throughout this paper, we have discussed the uncertainties inherent in
our method and assumptions. Here we briefly review these: (a) the
dispersion of the Gaussian distribution is varied through the range
$\sigma = 1.0\pm0.2$ mag, consistent with observations; (b) the
faint-end slope of the Schechter function is varied through the range
$\alpha = -1.41\pm0.05$, consistent with this and other analyses; (c)
Poisson variations in the LF of faint background objects are accounted
for; (d) possible overestimates of the EDGR which could result from
the inclusion of late-type dIr galaxies in the sample are
counterbalanced by a similar fraction of late-type giants also present
in the sample (both late-type dwarfs and giants are present only at a
very low level); and (e) the completeness function is involved in the
correction of both the program field and the control field LF, though
at such a low level that plausible variations in it would have only a
small effect.  Of these factors, (a) and (b) dominate the uncertainty
estimates on our estimate of the Coma EDGR.

Of immediate interest is whether our computed EDGR for the Coma
cluster exceeds that measured for Virgo, and thus if the EDGR-richness
correlation continues to this still-richer environment.  In the top
panel of Figure \ref{fig3}, we compare our derived EDGR values for the
Coma cluster (i.e., solid circles, from Table \ref{tbl-2}) with those
computed for Virgo (i.e., open circles; FS91\markcite{fs91}).  These
two sets of values are entirely consistent at all four limiting $R$
magnitudes. This reflects the fact that the faint-end slopes of the
galaxy LF are similar in both clusters.  Note that to a limit of $R =
18.6$ mag, TG93\markcite{tg93} measured an EDGR of 1.52, slightly
lower than (but formally consistent with) our value. While the primary
difference between our values results from the the modeling of the
log-normal distribution (TG93 fix $\sigma = 1.2$ mag), it must also be
due to the inclusion of lower-surface-brightness dE galaxies in our
deeper CCD sample (S95\markcite{s95a}; SH96\markcite{s95b}).

\placefigure{fig3}

In the bottom panel of Figure \ref{fig3}, we plot the EDGR (calculated
to a limit of $R = 21.6$ mag) for five galaxy groups taken from
FS91\markcite{fs91}.  To place Coma on this figure at the correct
richness, we adopt N(E+S0) = 291 (TG93\markcite{tg93}) as a measure of
the total number of giants in the entire Coma cluster.  Adopting this
value places Coma as easily the richest galaxy cluster in this sample,
in agreement with the Abell richness class rankings.  Then, we use a
Coma EDGR equal to 9.41$\pm$2.14 (Table \ref{tbl-2}) to scale the
number of measured dwarf galaxies for the cluster core to the Coma
cluster as a whole.  In this manner, we can put our derived EDGR for
the Coma cluster on the richness scale used by FS91\markcite{fs91} and
Vader and Sandage (1991)\markcite{vs91} (i.e.  $\log N_{\rm g}$).

In this figure, the solid line has unit slope, representing a constant
EDGR with increasing cluster richness: the Fornax, Virgo and Coma
clusters clearly lie well above this line, while the Leo, Dorado, and
NGC 1400 groups fall below.  Then from Figure \ref{fig3} and Table 2,
{\em the EDGR value for the Coma cluster is consistent with that found
for the Virgo cluster, within our calculated uncertainties}, and it is
clearly lower than would have been expected if the EDGR was to keep
increasing monotonically with cluster richness.

\section{DISCUSSION}

Rich galaxy clusters exhibit considerable variation within individual
richness classes, including properties such as: (i) the velocity
dispersion of the member galaxies; (ii) the Bautz-Morgan class, which
measures the magnitude difference between the first- and second-
ranked cluster galaxies; (iii) the presence or absence of a central cD
galaxy, (iv) X-ray emission; (v) evidence for significant
substructure.  Substructuring in particular must clearly be an
important factor in determining the early-type dwarf-to-giant ratio.
For example, if two or more richness-class 1 galaxy clusters merge to
form a richer cluster, then to first order the total number of dwarf
and giant galaxies will be conserved, and the EDGR for the new
(richer) cluster will mimic those for the original subclusters, rather
than being larger as suggested by the basic correlation shown in
Figure 3.  Thus the simple EDGR-richness correlation must really
represent an EDGR-richness-subcluster correlation (if not other
parameters too).  We are not suggesting that substructure replaces the
dependence on richness: there remains an EDGR-richness correlation for
{\em original} groups and clusters. However, substructure can explain
why the EDGR for {\em merger-product} clusters can mimic the EDGR of
the less-rich subgroups, instead of being as rich as predicted by the
EDGR-richness correlation.

For the case of Coma and Virgo, our results are consistent with a
scenario in which, identifiable subgroups in these clusters are richer
than the sparse clusters in the FS91 sample of poor groups, and are
consistent with the Coma cluster being the result of a significant
number of merger events (e.g., Fitchett \& Webster \markcite{fw87}
1987; Escalera, Slezak \& Mazure \markcite{esm92} 1992; Davis \&
Mushotzky \markcite{dm93} 1993; Mohr, Fabricant \& Geller
\markcite{mfg93} 1993; White, Briel \& Henry \markcite{wbh93} 1993;
West, Jones \& Forman \markcite{wjf94} 1995;  Colless \& Dunn 1996).
We postulate that to find an EDGR value which is clearly larger than
the Coma and Virgo values, it will be necessary to survey a cluster
environment which is rich {\it by birth}, rather than by subsequent
growth.  To turn this statement around, we suggest that a clear
signature of the very richest, densest galaxy environments (and by
extension, the most massive cluster potential wells in the early
universe) should be the presence of a significantly higher EDGR than
Coma or Virgo. Thus, for the EDGR-richness correlation to be relevant
to the formation and evolution of dE galaxies and the galaxy cluster
itself, we must disentangle the effects of richness and substructure.

Finally, SH96\markcite{s95b} discusses two additional factors affecting
the EDGR for rich galaxy clusters: (i) variation of the EDGR with
clustercentric radius, due to the destruction of faint dEs in the
dense core, and (ii) a predicted increase in the EDGR with cluster
X-ray luminosity, resulting from a confinement pressure exerted by
dense intracluster gas on proto-dE galaxies; i.e., more of the young
dwarf elliptical galaxies in gas-poor clusters should have expelled
their metal-rich gas and subsequently faded (cf., Babul \& Rees 
\markcite{br92} 1992).  Further surveys of other, more distant rich 
galaxy clusters to trace out these effects would be of great interest.

\acknowledgments

This research was supported in part by the Natural Sciences and
Engineering Research Council of Canada, through a grant to W.E.H., and
by the Ontario Ministry of Colleges and Universities, through an Ontario
Graduate Scholarship to J.S. We would like to thank an anonymous
referee for helpful comments, and to acknowledge the assistance of
Stephen Holland with the observations.

\clearpage

\begin{center}
\begin{deluxetable}{llclrccc}
\footnotesize
\tablecaption{Fits of the Gaussian-plus-Schechter model to the galaxy luminosity 
function. \label{tbl-1}}
\tablewidth{0pt}
\tablehead{\colhead{Fit} & \multicolumn{3}{c}{Gaussian} & \multicolumn{3}{c}
{Schechter} & \colhead{$\chi^{2}/\nu$} \nl
            & \colhead{$\kappa_1$} & \colhead{$m^0$} & 
\colhead{$\sigma$} & \colhead{$\kappa_2$} & \colhead{$m^{\star}$}
& \colhead{$\alpha$}   & }
\startdata
 1 & 15.30$\pm$9 & 14.53$\pm$0.01 & 0.80 & 14.74$\pm$19 & 15.61$\pm$0.37 & 
--1.41 & 0.87 \nl
 1a& 14.29$\pm$13& 14.48$\pm$0.01 & 0.80 & ~9.89$\pm$21 & 15.10$\pm$0.91 & 
--1.46 & 0.85 \nl
 1b& 16.01$\pm$7 & 14.57$\pm$0.01 & 0.80 & 19.95$\pm$20 & 15.97$\pm$0.21 & 
--1.36 & 0.93 \nl
 2 & 15.18$\pm$7 & 14.66$\pm$0.01 & 0.90 & 17.30$\pm$24 & 15.96$\pm$0.31 & 
--1.41 & 0.86 \nl
 3 & 14.82$\pm$6 & 14.78$\pm$0.01 & 1.00 & 19.58$\pm$29 & 16.23$\pm$0.28 & 
--1.41 & 0.90 \nl
 4 & 14.40$\pm$6 & 14.90$\pm$0.01 & 1.10 & 21.57$\pm$35 & 16.44$\pm$0.26 & 
--1.41 & 0.98 \nl
 5 & 14.04$\pm$5 & 15.04$\pm$0.01 & 1.20 & 23.33$\pm$41 & 16.61$\pm$0.26 & 
--1.41 & 1.06 \nl
5a & 13.72$\pm$6 & 15.01$\pm$0.01 & 1.20 & 18.08$\pm$40 & 16.37$\pm$0.41 & 
--1.46 & 1.09 \nl
5b & 14.30$\pm$5 & 15.05$\pm$0.01 & 1.20 & 28.98$\pm$44 & 16.82$\pm$0.18 & 
--1.36 & 1.04 \nl
\enddata
\end{deluxetable}
\end{center}

\clearpage

\begin{center}
\begin{deluxetable}{crrrrrrrrrcc}
\footnotesize
\tablecaption{N(dE+dE,N+dS0) and the EDGR for different function 
fits and limiting magnitudes. \label{tbl-2}}
\tablewidth{0pt}
\tablehead{
\colhead{$R$-Mag}  & \multicolumn{9}{c}{N(dE+dE,N+dS0)}  & 
\colhead{Coma}   &  \colhead{Virgo} \nl
Limit & Fit \#1 & 1a & 1b & 2 & 3 & 4 & 5 & 5a & 5b & EDGR & EDGR}
\startdata
18.6  & ~66.9  & ~67.8 & ~66.5 & ~62.3  & ~58.2 & ~54.7 & ~51.7 & ~51.3 & ~51.6 
& 1.80$\pm$0.58 & 2.12   \nl
19.6  & 120.1  & 120.7 & 120.0 & 116.2  & 112.4 & 109.0 & 106.0 & 104.7 & 106.4 
& 3.38$\pm$0.86 & 3.61   \nl
20.6  & 199.7  & 202.8 & 197.0 & 197.5  & 195.0 & 192.5 & 190.2 & 190.4 & 188.6 
& 5.80$\pm$1.33 & 5.77   \nl
21.6  & 316.8  & 328.9 & 305.9 & 317.6  & 317.5 & 316.9 & 316.0 & 323.8 & 306.6 
& 9.41$\pm$2.14 & 9.31   \nl
\enddata
\tablenotetext{1}{N(E+S0) is calculated over the range $12.5 \leq R \leq 18.5$
mag, and equals 30.51, 28.46, 31.95, 33.96, 36.72, 39.11, 41.42, 40.44 and 
42.20 for Fits 1 through 5b (as above).}
\tablenotetext{2}{The Virgo EDGR values are taken from Ferguson \& Sandage
\markcite{fs91} (1991).}
\end{deluxetable}
\end{center}

\clearpage

\figcaption[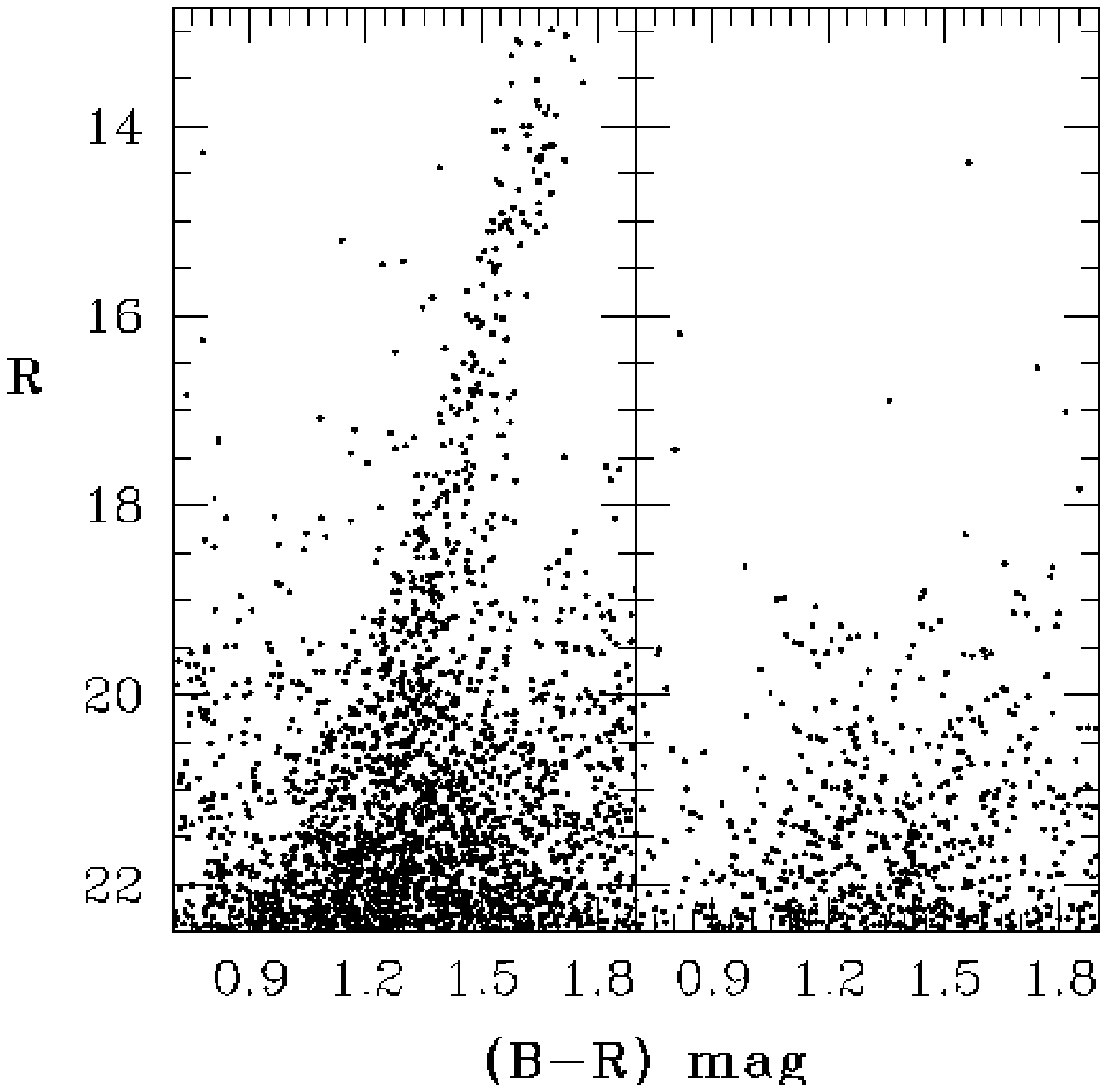]{Color-magnitude diagrams (CMDs) for detected objects
on the cluster fields (left panel) and on the control field, which is
2 degrees from the cluster center (right panel).  The cluster galaxies
(dwarf and giants) are immediately obvious as a sequence of objects
which are restricted to a narrow range in $(B-R)$ color, and which are
not present on the control field CMD. \label{fig1}}

\figcaption[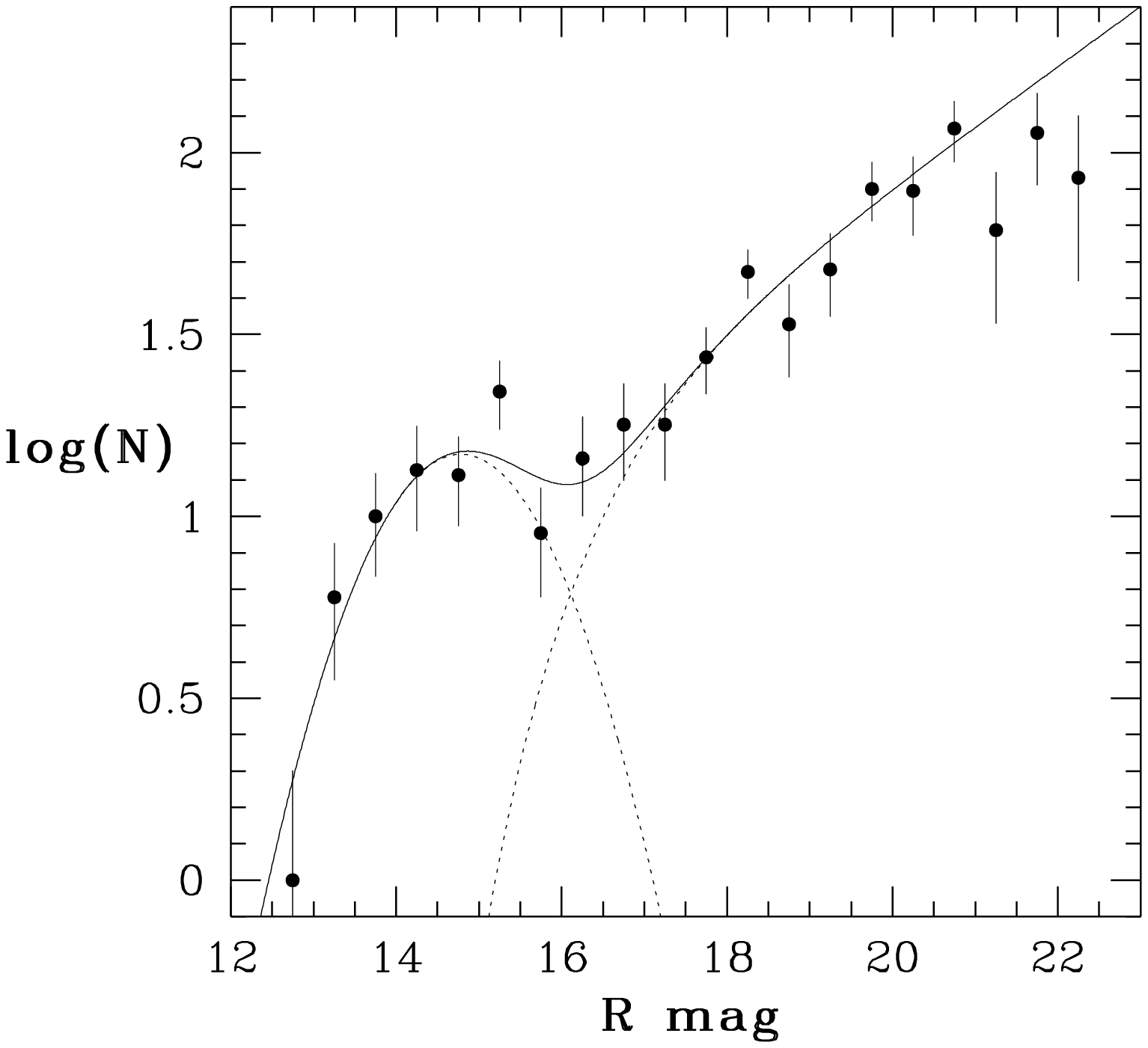]{The composite galaxy luminosity function (LF) 
for the Coma cluster core. Solid dots represent the total number of
Coma galaxies (after correction for incompleteness and background
subtraction) per half-magnitude bin.  The solid line shows the model
fit, the sum of two separate contributions (dashed lines) from the
giants (i.e., a log-normal distribution) and dwarf ellipticals (i.e.,
a Schechter function).  In this plot we illustrate model fit 3, for which
the Gaussian distribution has a dispersion of $\sigma = 1.0$ mag; the
other relevant parameters are given in Tables 1 and 2. The three
faintest points were not included in the fit, as they adversely affect
normalization over the region of interest. \label{fig2}}

\figcaption[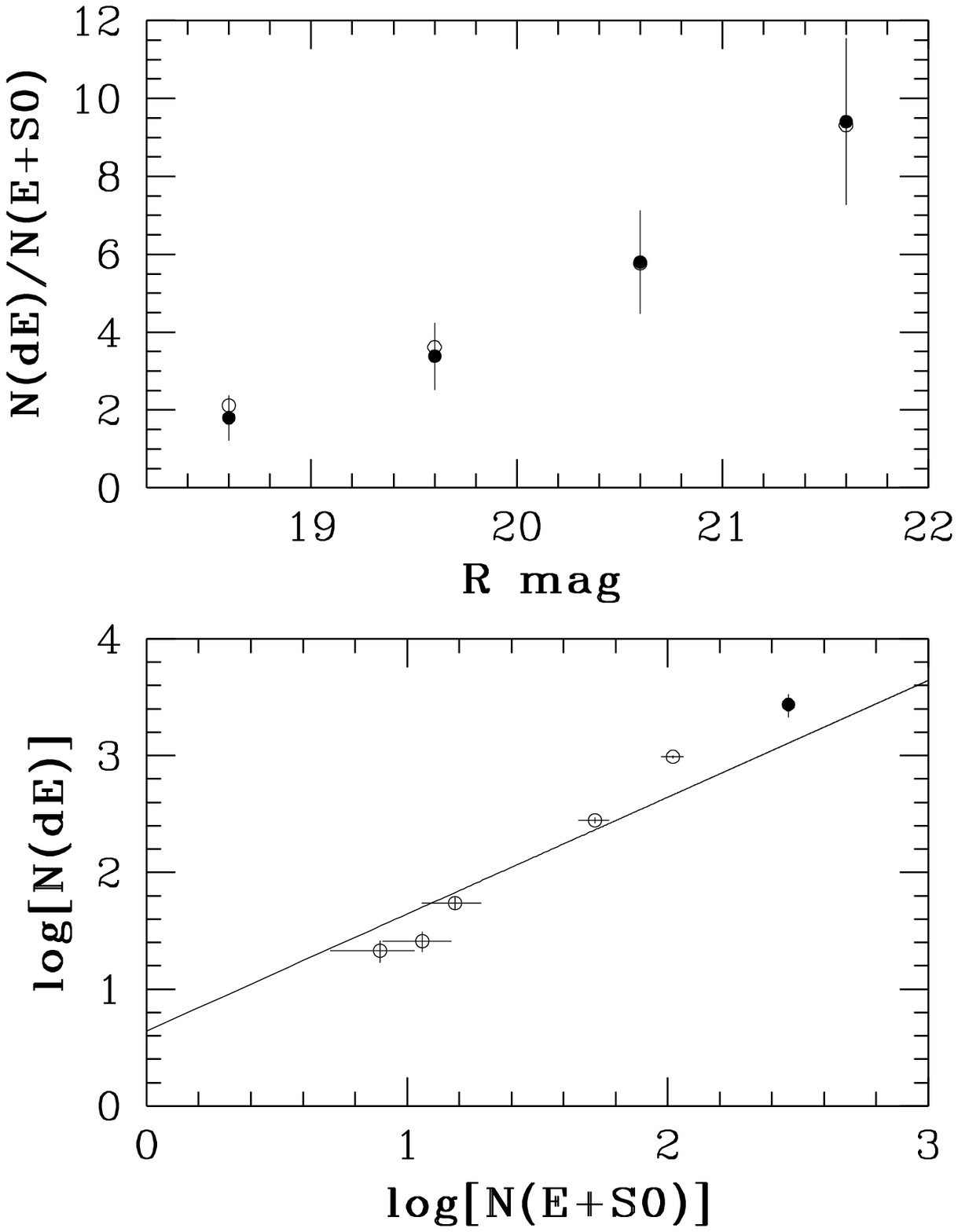]{The early-type dwarf-to-giant ratio (EDGR) for 
the Coma cluster. In both panels, the solid circle corresponds to our
measured EDGR for the Coma cluster. In the top panel it is clear that
the Coma and Virgo EDGRs are identical within the uncertainties (those
shown are for our computed values), for the four different magnitude
limits used in the numerical integration of the Schechter function. In
the lower panel we plot the number of early-type dwarf galaxies (to a
$R= 21.6$ mag limit) versus the number of early-type giant galaxies.
From left to right, the open circles represent Leo, Dorado, NGC 1400,
Fornax and Virgo (from FS91), while the solid dot is the Coma value
from our study, and the solid line has a slope of unity. \label{fig3}}

\clearpage

\plotone{fig1.eps}

\clearpage

\plotone{fig2.eps}

\clearpage

\plotone{fig3.eps}
\end{document}